\documentclass[3p,times]{elsarticle}

\usepackage{ecrc}
\usepackage{axodraw4j}

\volume{00}

\firstpage{1}

\journalname{Nuclear Physics A}

\runauth{David d'Enterria, Alexander Snigirev}

\jid{npa}

\jnltitlelogo{Nuclear Physics A}

\usepackage{amssymb}

\biboptions{square,comma,numbers,sort&compress}

\usepackage[figuresright]{rotating}

\usepackage{multirow}

\newcommand{\sqrtsnn}{\sqrt{s_{_{\mbox{\rm \tiny{NN}}}}}}
\newcommand{\alphaS}{\alpha_{\rm s}}

\newcommand{\pp}{{\rm{p-p}}}

\newcommand{\pA}{{\rm{p-A}}}
\newcommand{\pN}{{\rm{p-N}}}
\newcommand{\pPb}{{\rm{p-Pb}}}
\newcommand{\NN}{{\rm{N-N}}}
\newcommand{\AaAa}{{\rm{A-A}}}
\newcommand{\PbPb}{{\rm{Pb-Pb}}}
\newcommand{\TpA}{\rm T_{_{\rm pA}}}

\newcommand{\TpAsq}{\rm T^{2}_{_{\rm pA}}}

\newcommand{\dtwor}{{\rm{d^2r}}}

\newcommand{\mcfm}{{\sc mcfm}}

\newcommand{\sigmaDPS}{\sigma^{{\rm {\tiny DPS}}}}

\newcommand{\sigmaSPS}{\sigma^{{\rm {\tiny SPS}}}}
\newcommand{\sigmaeff}{\sigma_{\rm eff}}
\newcommand{\sigmaeffpp}{\sigma_{_{\rm eff,pp}}}
\newcommand{\sigmaeffpA}{\sigma_{_{\rm eff,pA}}}
\newcommand{\sigmaeffAA}{\sigma_{_{\rm eff,AA}}}
\newcommand{\sigmaDPSjpsijpsi}{\sigma^{{\rm {\tiny DPS}}}_{{\jpsi\jpsi}}}

\newcommand{\NDPS}{\rm N^{{\rm {\tiny DPS}}}}

\newcommand{\jpsi}{J/\psi}

\def\order#1{\mathcal{O}{(#1)}}

\begin{document}

\begin{frontmatter}



\dochead{}

\title{Double-parton scattering cross sections in proton-nucleus\\ and nucleus-nucleus collisions at the LHC}


\author[cern]{\underline{David d'Enterria}} 
\address[cern]{CERN, PH Department, 1211 Geneva, Switzerland}
\author[msu]{Alexander~M.~Snigirev}
\address[msu]{Skobeltsyn Institute of Nuclear Physics, Moscow State University, 119991 Moscow, Russia}

\begin{abstract}
Simple generic expressions to compute double-parton scattering (DPS) cross sections in high-energy 
proton-nucleus and nucleus-nucleus collisions, as a function of the corresponding single-parton 
cross sections, are presented. Estimates of DPS contributions are studied for two specific processes at 
LHC energies: (i) same-sign W-boson pair production in \pPb, and (ii) double-$\jpsi$ production in \PbPb, 
using NLO predictions with nuclear parton densities for the corresponding single-parton cross sections. 
The expected DPS cross sections and event rates after typical acceptance and efficiency losses are also given
for other processes involving $\jpsi$ and W,Z gauge bosons in \pPb\ and \PbPb\ collisions at the LHC.
\end{abstract}


\end{frontmatter}


\section{Introduction}
\label{sec:intro}

Hadrons (protons, nuclei) are composite particles with a finite transverse size, whose internal parton density
rises rapidly with increasing colliding energies. Such a fact leads naturally to a large number of
multiple parton interactions (MPI) occurring in each single proton-proton (\pp), proton-nucleus (\pA), and
nucleus-nucleus (\AaAa) collision at LHC energies~\cite{Bartalini:2010su,Bartalini:2011jp}. Many basic \pp\
event properties --such as the distributions of hadron multiplicities in ``minimum bias'' collisions and the
underlying event activity in hard scattering interactions-- can only be reproduced by Monte Carlo 
generators by including MPI at semi-hard scales of order $\order{1-3\rm~GeV}$, modeled through an
impact-parameter description of the transverse parton profile of the colliding protons. The experimental
evidence for the occurrence of double-parton scatterings (DPS) producing in the same collision 
two independently-identified particles at harder scales, $\order{3-100\rm~GeV}$, is however scarcer. 
The study of DPS processes provides valuable information on the transverse distribution of partons in
hadrons~\cite{Diehl:2011yj}, on multi-parton correlations in the hadronic wave
functions~\cite{Calucci:2010wg}, as well as on backgrounds for new physics signals. The importance  
of DPS in \pA\ and \AaAa\ collisions has been quantitatively highlighted for the first time in
Refs.~\cite{d'Enterria:2012qx,d'Enterria:2013ck}. A summary of those results is presented below together with
new estimates of DPS cross sections for various double hard processes in nuclear collisions at the LHC.\\ 

In a generic model-independent way, one can write the DPS cross section in \pp\
collisions as the product of the single-parton scattering (SPS) cross sections, $\sigmaSPS$
--computable perturbatively to a given order in the strong coupling $\alphaS$ by convoluting the partonic
subprocess cross sections with the corresponding parton distribution functions (PDF)-- normalized by an
effective cross section $\sigmaeffpp$ characterizing the transverse area of the hard partonic interactions:
\begin{equation} 
\sigmaDPS_{(pp\to a b)} = \left(\frac{m}{2}\right) \frac{\sigmaSPS_{(pp\to a)} \cdot \sigmaSPS_{(pp\to b)}}{\sigmaeffpp}\,,
\label{eq:sigmaDPSpp}
\end{equation}
where the combinatorial factor $m/2$ accounts for indistinguishable ($m=1$) and distinguishable ($m=2$) final-states.
A numerical value $\sigmaeffpp\approx$~15~mb has been obtained in \pp\ at the LHC from empirical fits to W+dijets
distributions sensitive to DPS contributions~\cite{Aad:2013bjm,Chatrchyan:2013xxa}.
One can identify $\sigmaeffpp$ with the inverse of the proton overlap-function squared: 
$\sigmaeffpp = \left[ \int d^2b \, t^2({\bf b})\right]^{-1}$ under the two following 
assumptions: (i) the double-PDF can be decomposed into longitudinal  
and transverse components, with the latter expressed in terms of the overlap function 
$t({\bf b}) = \int f({\bf b_1}) f({\bf b_1 -b})d^2b_1 $ for a given parton transverse thickness function 
$f({\bf b})$ representing the effective transverse overlap area of partonic interactions that produce the DPS
process, and (ii) the longitudinal component reduces to the ``diagonal'' product of two independent
single-PDF. The fact that the measured $\sigmaeffpp$ is about a factor of two smaller than estimates based on naive
geometric descriptions of the proton~\cite{Abe:1997xk}, and ascertaining the evolution of $\sigmaeffpp$ with 
collision energy, remain two important open issues in DPS studies.\\

In \pA\ collisions, the parton flux is enhanced by the number A of nucleons in the nucleus and the SPS cross
section is simply that of proton-nucleon (\pN) collisions (accounting for shadowing effects in the 
nuclear PDF) 
scaled by A~\cite{d'Enterria:2003qs}, 
whereas the DPS cross sections is further enhanced due to interactions where the two partons of the nucleus belong
to the same nucleon and to two different nucleons~\cite{Strikman:2001gz}. 
The corresponding DPS cross section 
reads~\cite{d'Enterria:2012qx}:
\begin{eqnarray} 
\sigmaDPS_{pA\to a b} = \left(\frac{m}{2}\right) \frac{\sigmaSPS_{pN \to a} \cdot \sigmaSPS_{pN \to b}}{\sigmaeffpA}\,,
\mbox{ with } \; \sigmaeffpA = \frac{\sigmaeffpp}{A+\sigmaeffpp\,\rm{F}_{pA}} \approx 22.6 
\mbox{ $\mu$b},
\label{eq:sigmapADPS}
\end{eqnarray} 
where $\rm{F}_{pA} = \frac{A-1}{A} \int \TpAsq({\bf r})\,\dtwor$ with  the standard Glauber nuclear thickness
function $\TpA({\bf r})$~\cite{d'Enterria:2003qs}, and the last numerical equality holds for \pPb\ using 
A~=~208, $\sigmaeffpp =$~15~mb and $\rm{F}_{pA}$~=~30.4~mb$^{-1}$. The DPS cross sections
in \pPb\ are thus enhanced by a factor of $\sigmaeffpp/\sigmaeffpA \approx 3\,A \approx$~600 compared to \pp.
Exploiting such a large expected DPS signal allows one
to determine the value of $\sigmaeffpp$ independently of other \pp\ measurements,
given that the parameter F$_{pA}$ depends on the comparatively better known transverse density profile of nuclei.\\

\begin{figure}[hbtp!]
  \centering
  \includegraphics[width=0.85\columnwidth]{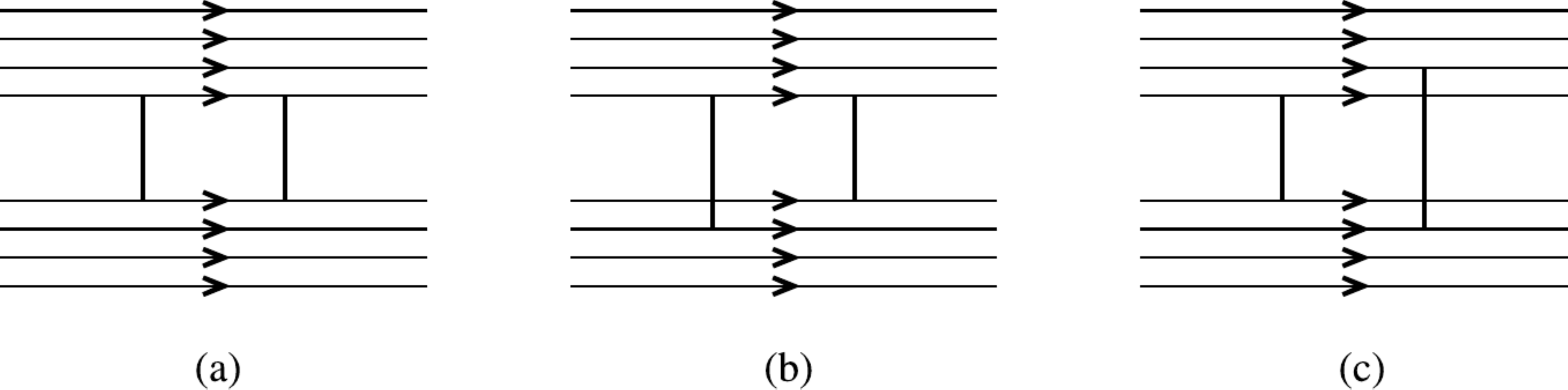}
\caption{Schematic DPS contributions in \AaAa\ collisions: 
(a) The two colliding partons belong to the same pair of nucleons, 
(b) partons from one nucleon in one nucleus collide with partons from two different nucleons in the other nucleus, and 
(c) the two colliding partons belong to two different nucleons from both nuclei.
\label{fig:diags}}
\end{figure}

In the \AaAa\ case, the single-parton cross section is that of \pp\ (again, modulo nuclear PDF shadowing corrections)
scaled by A$^2$, and the DPS one is the sum of the three terms shown in Fig.~\ref{fig:diags} which
result in the final expression~\cite{d'Enterria:2013ck}:
\begin{eqnarray} 
\sigmaDPS_{(AA\to a b)} = \left(\frac{m}{2}\right) \frac{\sigmaSPS_{(NN \to a)} \cdot \sigmaSPS_{(NN \to b)}}{\sigmaeffAA},
\mbox{ with } \; \sigma_{\rm eff,AA} = \frac{1}{A^2\left[\sigmaeffpp^{-1}+\frac{2}{A}\,\rm{T}_{AA}(0)\,+\,\frac{1}{2}\,\rm{T}_{\rm AA}(0)\right]} \approx 1.5 \mbox{ nb} \,.
\label{eq:sigmaAADPS}
\end{eqnarray}
where $\rm{T}_{AA}({\bf b})$ is the nuclear overlap function~\cite{d'Enterria:2003qs}  
amounting to $\rm{T}_{AA}(0)$~=~30.4~mb$^{-1}$ for head-on \PbPb\ interactions.
The relative contributions of the three terms in the denominator, corresponding to the diagrams of
Fig.~\ref{fig:diags}, are approximately 1:4:200. Whereas the single-parton cross sections in
\PbPb\ are enhanced by a factor of A$^2~\simeq~4\cdot 10^4$ compared to that in \pp, the corresponding
double-parton cross sections are enhanced by a much higher factor of  
$\sigmaeffpp\,/\sigma_{\rm eff,AA}\propto A^{3.3}/5 \simeq 9 \cdot 10^6$.
Pair-production of pQCD probes issuing from DPS represents thus an important feature of heavy-ion collisions at the
LHC and needs to be taken into account in any attempt to fully understand the event-by-event characteristics
of any yield suppression and/or enhancement observed in \PbPb\ compared to \pp\ data.\\

\section{Double parton scatterings in p-Pb $\rightarrow$ W$^+$W$^+$ at 8.8 TeV, and in Pb-Pb $\rightarrow\jpsi\jpsi$ at 5.5 TeV}
\label{sec:ppb}

\begin{figure}[hbtp!]
  \centering
  \includegraphics[width=0.49\columnwidth,height=8.4cm]{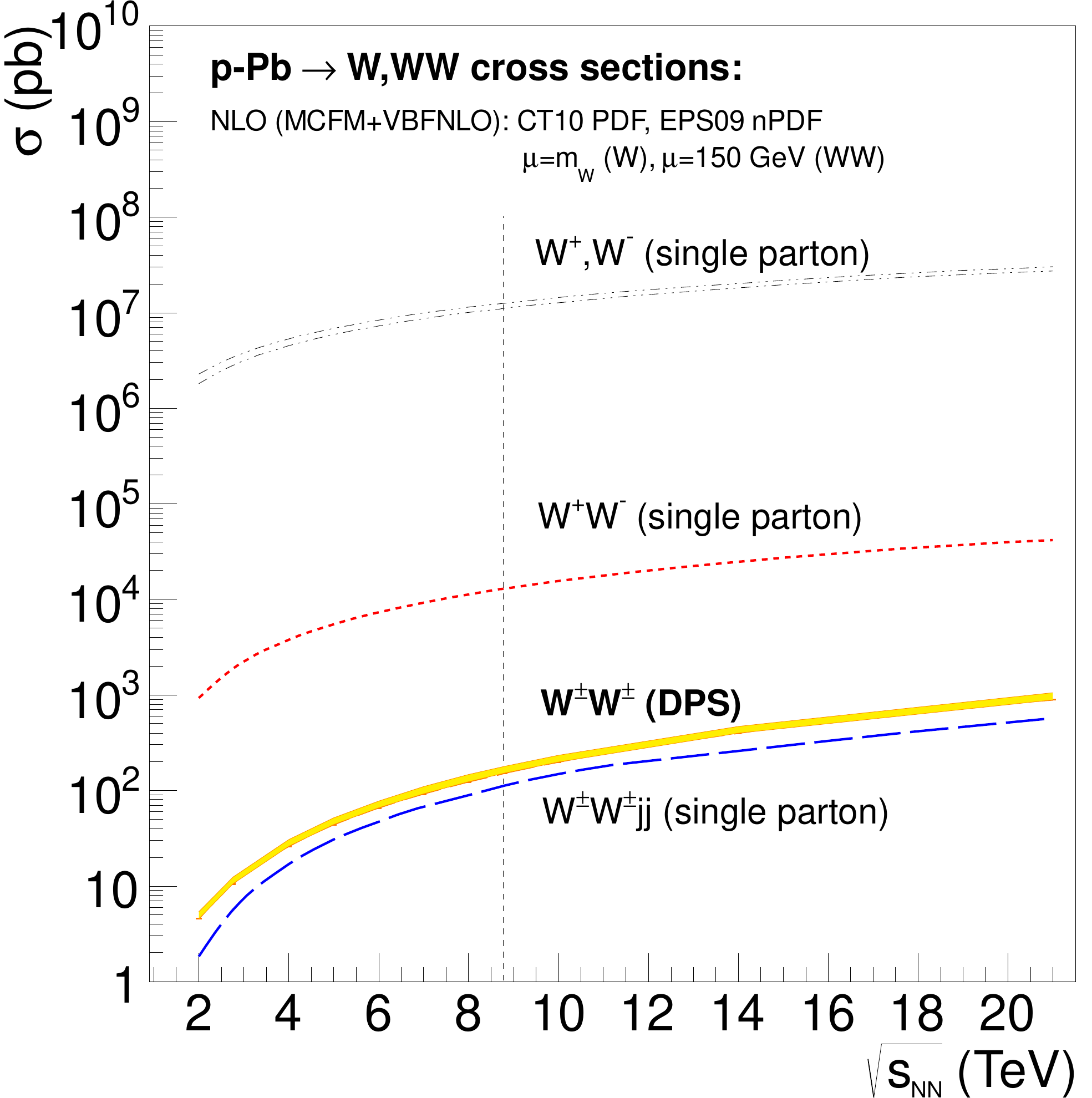} 
  \hspace{0.2cm}
  \includegraphics[width=0.47\columnwidth,height=8.2cm]{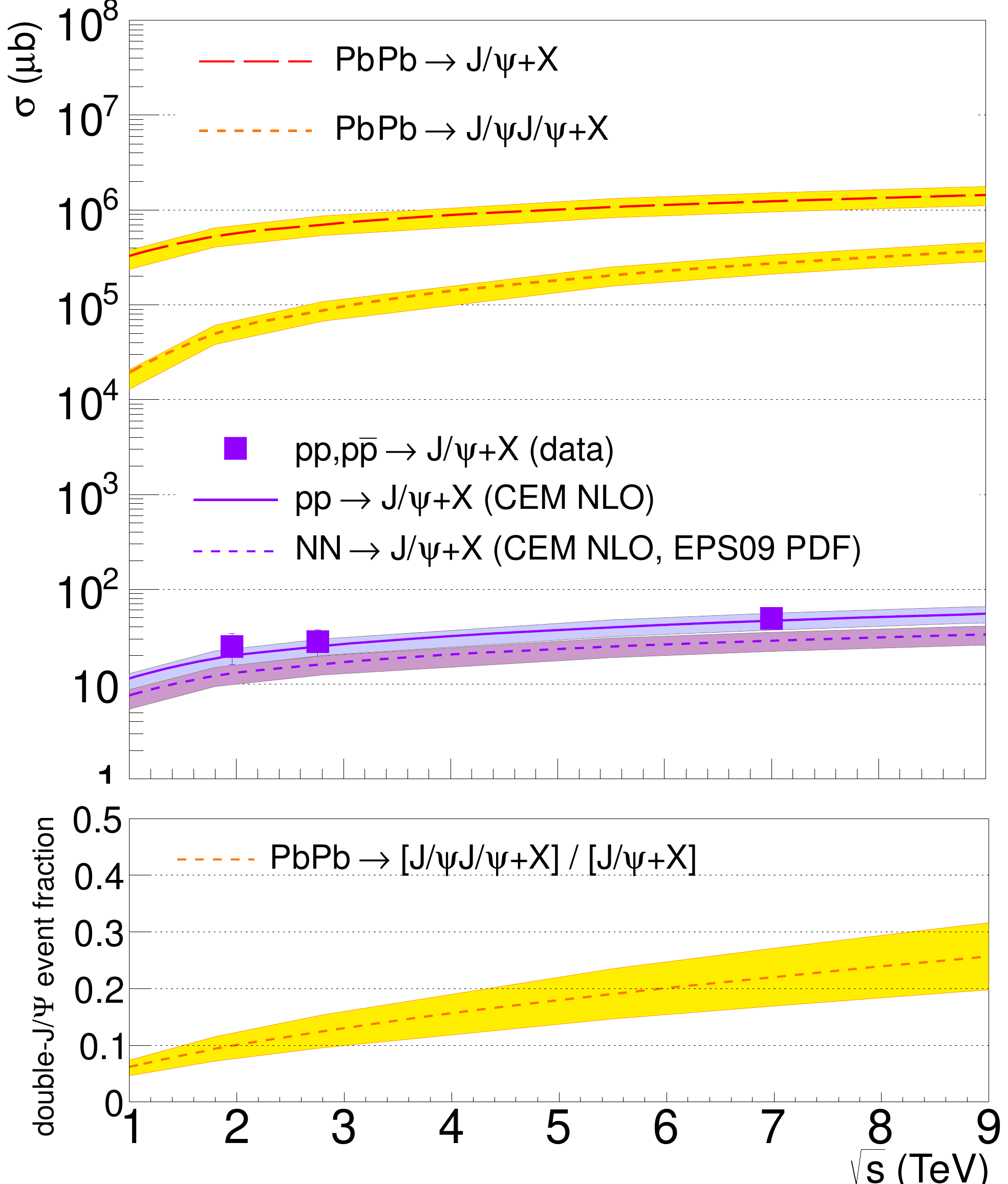}
  \caption{Cross sections as a function of c.m. energy
  for single-W 
  and W-pair boson(s) in \pPb\ for single-parton and double-parton scatterings (left)~\cite{d'Enterria:2012qx};
  and for prompt-$\jpsi$ production in \pp, \NN, and \PbPb\ collisions and for
  double-parton $\jpsi\jpsi$ in \PbPb\ (right)~\cite{d'Enterria:2013ck}.} 
  \label{fig:sigmaDPS_vs_sqrts}
\end{figure}

The production of like-sign WW production --whose cross section has small theoretical uncertainties and a
characteristic final-state with same-sign leptons plus (large) missing transverse energy 
from the undetected neutrinos-- has no SPS backgrounds at the same order in $\alphaS$, and has been proposed
since long as a DPS ``smoking gun'' in \pp\ collisions~\cite{Kulesza:1999zh}. 
The DPS signal in \pPb\ collisions, $\sigmaDPS_{pPb\to WW}$, has been computed via Eqs.~(\ref{eq:sigmapADPS}) 
with NLO single-parton W cross section $\sigmaSPS_{pN \to W}$ using \mcfm~6.2~\cite{Campbell:2011bn} 
with CT10 proton~\cite{Lai:2010vv} and EPS09 nuclear~\cite{Eskola:2009uj} PDF and 
theoretical scales $\mu = \mu_F = \mu_R$~=~$m_W$.
Figure~\ref{fig:sigmaDPS_vs_sqrts} (left) shows the total cross sections for all relevant 
processes in the range of nucleon-nucleon c.m. energies $\sqrtsnn \approx$~~2--20~TeV. 
At the nominal 8.8~TeV, the same-sign WW DPS cross section is $\sigmaDPS_{pPb\to WW}\approx$~150~pb (yellow
thick curve), i.e. a factor of 1.5 times higher than the SPS background, $\sigmaSPS_{pPb\to WWjj}$, 
obtained adding the QCD and electroweak cross sections for the production of W$^+$W$^+$ (W$^-$W$^-$)
plus 2 jets pairs (lowest dashed curve). Accounting for the leptonic decay ratios and applying
standard ATLAS/CMS 
acceptance and reconstruction cuts, one expects about 10 DPS same-sign WW
events in 2~pb$^{-1}$ integrated luminosity~\cite{d'Enterria:2012qx}.


The double-$\jpsi$ DPS cross section in \PbPb\ has been computed with
Eqs.~(\ref{eq:sigmaAADPS}) using the colour evaporation model (CEM)~\cite{Vogt:2012vr}, which agrees
well with the Tevatron and LHC data (squares in Fig.~\ref{fig:sigmaDPS_vs_sqrts} right),
with EPS09 nuclear PDF 
for the SPS $\jpsi$ cross section, $\sigmaSPS_{(NN \to \jpsi\,X)}$.
The two top curves in 
Fig.~\ref{fig:sigmaDPS_vs_sqrts} (right) show the 
single-$\jpsi$ (dashes) and double-$\jpsi$ (dots) cross sections in \PbPb, whereas their ratio
is plotted in the bottom panel. At the nominal \PbPb\ energy of 5.5~TeV, single prompt-$\jpsi$ cross sections
amount to $\sim$1~b, and $\sim$20\% of such collisions are accompanied by the production of a second $\jpsi$
from a double parton interaction. The probability of $\jpsi$-$\jpsi$ DPS production increases rapidly with
centrality and at the lowest impact-parameters $\sim$35\% of the \PbPb$\,\to\jpsi+X$ collisions have a
second $\jpsi$ in the final state~\cite{d'Enterria:2013ck}. Accounting for decays, acceptance and efficiency
--which result in a $\sim$3$\cdot$10$^{-7}$ reduction factor in the ALICE (forward) and ATLAS/CMS
(central) rapidities-- the visible cross section is $d\sigmaDPSjpsijpsi/dy|_{y=0,2} \approx$~60~nb per
dilepton decay mode, i.e. about 240 double-$\jpsi$ events per unit-rapidity in the four
combinations of dielectron and dimuon channels in $\cal{L}_{\rm int}$~=~1~nb$^{-1}$, assuming
no in-medium $\jpsi$ suppression (accounting for it would reduce the yields by a factor of~2). 
These results show quantitatively that the observation of a $\jpsi$ pair in a given \PbPb\ event
should not be (wrongly) interpreted as indicative of $\jpsi$ production via $\rm c \bar{c}$ regeneration in
the dense medium created, as DPS are an important component of the total $\jpsi$
yield with or without final-state quark-gluon-plasma effects.  

\section{Compilation of double-parton scatterings in p-Pb and Pb-Pb }
\label{sec:compilation}

Table~\ref{tab:1} collects SPS and DPS cross sections involving $\jpsi$ and electroweak bosons in \pPb\ and \PbPb\
collisions at the LHC.
The SPS cross sections are computed at NLO (CEM for $\jpsi$ and \mcfm\ for W,Z), and the quoted visible DPS 
yields (for $\cal{L}_{\rm int}$~=~1~pb$^{-1}$ and 1~nb$^{-1}$)
take into account: BR($\jpsi$,W,Z)~=~6\%,~11\%,~3.4\% per dilepton decay; 
and (simplified) acceptance+efficiency losses: ${\cal A\times E}(\jpsi$,W,Z)~$\approx$~0.12 (per unit rapidity),
0.5, 0.5 (full rap.). These estimates indicate that DPS processes leading to
double-$\jpsi$, $\jpsi$+W, $\jpsi$+Z, and same-sign WW, are indeed observable in LHC heavy-ion runs. 
Other DPS processes like W+Z and Z+Z have lower visible cross sections and are not quoted.

\begin{table}[htbp]
\renewcommand{\arraystretch}{1.3}
\begin{tabular}{l|lccc|llccccc}\hline
$\sqrtsnn$ &  & $\sigmaSPS(\jpsi)$ & $\sigmaSPS$(W$^+$)  & $\sigmaSPS$(Z) & 
 & & $\jpsi\,\jpsi$ & $\jpsi$+W & $\jpsi$+Z & ss\,WW \\\hline
5.5 TeV & \NN & 25 $\mu$b & 30 nb & 20 nb & \multirow{2}{*}{\PbPb} & $\sigmaDPS$ & 200 mb & 500 $\mu$b & 330 $\mu$b & 630 nb \\
 & &  &  &  &  & $\NDPS$(1 nb$^{-1}$) & $\sim$240 & $\sim$80 & $\sim$10 & $\sim$20 \\\hline
8.8 TeV & p-N& 45 $\mu$b & 60 nb & 35 nb & \multirow{2}{*}{\pPb} & $\sigmaDPS$ & 45 $\mu$b & 120 nb & 70 nb & 140 pb \\
 & &  &  &  &  & $\NDPS$(1 pb$^{-1}$) & $\sim$60 & $\sim$15 & $\sim$2 & $\sim$5 \\\hline
\end{tabular}
\caption{Production cross sections of prompt-$\jpsi$, W and Z bosons in single-parton-scatterings (SPS) in
  nucleon-nucleon (\NN) and proton-nucleon (p-N) collisions, and of $\jpsi$-$\jpsi$, $\jpsi$+W, $\jpsi$+Z, and
  same-sign WW in double-parton-scatterings (DPS) in \PbPb\ and \pPb\ at LHC energies. 
The corresponding approximate DPS yields (after dilepton decays and acceptance losses) are also given for 1~nb$^{-1}$ and
1~pb$^{-1}$ respectively.} 
\label{tab:1}
\vspace{-0.5cm}
\end{table}

\section{Summary}

Simple generic expressions to compute the double-parton cross sections in high-energy \pA\ and \AaAa\
collisions have been derived from the corresponding single-parton cross sections. The larger transverse parton
density in nuclei results in significantly enhanced cross sections for many DPS processes compared to \pp\ collisions.
Estimates of DPS contributions for (i) same-sign W pair production in \pPb\ and (ii) double-$\jpsi$ production in
\PbPb, have been obtained at LHC energies. The first process can help determine the effective $\sigmaeff$
parameter characterising the transverse parton distribution in the nucleon, and the second one
provides interesting insights on the event-by-event dynamics of $\jpsi$ production in \PbPb\ collisions. 
Many other DPS processes in \pPb\ and \PbPb\ (such as $\jpsi$+W, $\jpsi$+Z) have large
visible cross sections and are open to study at the LHC. 





\bibliographystyle{elsarticle-num}
\bibliography{dde_proceeds_hp13}







\end{document}